# Current transport and thermoelectric properties of very high power factor $Fe_3O_4$ / $SiO_2$ / *p*-type Si (001) devices

M. Zervos[1], Z. Viskadourakis[2,†], G. Athanasopoulos[1], R. Flores[3], O. Conde[3], and J. Giapintzakis[1,*]

1. Nanotechnology Research Center (NRC) and Department of Mechanical and Manufacturing Engineering, School of Engineering, University of Cyprus, 75 Kallipoleos, P.O. Box 20537, Nicosia 1678, Cyprus
2. Department of Materials Science and Technology, University of Crete, P.O. Box 2208, Heraklion, Crete 710 03, Greece
3. Department of Physics, University of Lisbon and ICEMS, Campo Grande Ed. C8, Lisbon 1749-016, Portugal

† Current address: Crete Center for Quantum Complexity and Nanotechnology, CCQCN University of Crete, P.O. Box 2208, Heraklion 71003, Greece

* Corresponding author e-mail address: giapintz@ucy.ac.cy

**Abstract**

The current transport and thermoelectric properties of $Fe_3O_4$ / $SiO_2$ / *p*-type Si(001) heterostructures with $Fe_3O_4$ thicknesses of 150, 200, and 350 nm have been investigated between 100 and 300 K. We observe a sharp drop of the in-plane resistivity at 200K due to the onset of conduction along the Si / $SiO_2$ interface related to tunneling of electrons from the $Fe_3O_4$ into the accumulation layer of holes at the Si / $SiO_2$ interface, whose existence was confirmed by capacitance-voltage measurements and a two band analysis of the Hall effect. This is accompanied by a large increase of the Seebeck coefficient reaching +1000 $\mu$V/K at 300K that is related to holes in the *p*-type Si(001) and gives a power factor of 70 mW/K$^2$m when the $Fe_3O_4$ layer thickness is reduced down to 150 nm. We show that most of the current flows in the $Fe_3O_4$ layer at 300 K, while the $Fe_3O_4$ / $SiO_2$ / *p*-type Si(001) heterostructures behave like tunneling p-n junctions in the transverse direction.

**Introduction**

Thermoelectric materials and devices have been investigated for many decades since the conversion of heat to electricity is important for the conservation of the environment. The conversion efficiency is described by the dimensionless *figure-of-merit*, $ZT = S^2 T / \rho \kappa$, where $S$ is the Seebeck coefficient or thermopower, $\rho$ the electrical resistivity, $T$ the temperature, and $\kappa$ the total thermal conductivity. A high ZT may be achieved by reducing $\kappa$ and/or by increasing the thermoelectric power factor $P = S^2 / \rho$; however, this is not easy due to the interdependence of $S$, $\rho$, and $\kappa$ since a reduction of $\rho$ is usually achieved by increasing the carrier density, which in turn reduces $S$ [1]. This constraint can be relaxed via the use of low dimensional semiconductors such as quantum wells, wires, or dots [2–6] after Hicks and Dresselhaus [7,8]. In addition, significant reductions in $\kappa$ have been attained via the use of nanostructured materials such as nanowires [9], while large power factors $P$ have been obtained by tuning the electronic properties [10]. For instance, a huge $P$, of 200 mW/K²m, was measured by Sun *et al.* [11] in $FeSb_2$ and shown to be electronic in origin. Recently, we investigated the lateral transport and thermoelectric properties of a 200 nm $Fe_3O_4$ / $SiO_2$ / *p*-type Si(001) heterostructure, which exhibited a large increase in the Seebeck coefficient up to +900 µV/K with increasing temperature and a concurrent sharp drop in the resistance giving a very high power factor of 25.5 mW/K²m. The sharp drop of resistance with increasing temperature was attributed to the onset of conduction along the Si / $SiO_2$ interface due to tunneling of electrons from the $Fe_3O_4$ layer into the accumulation layer of holes suggested to exist at the Si / $SiO_2$ interface as a consequence of the difference in the work functions of $Fe_3O_4$ and underlying *p*-type Si(001) [12].

Here, we have carried out a detailed investigation into the lateral and transverse current transport and thermoelectric properties of $Fe_3O_4$ / $SiO_2$ / *p*-type Si (001) heterostructures having $Fe_3O_4$ thicknesses of 150, 200, and 350 nm in order to gain deeper insight of the current flow distribution and origin of the high thermopower, which are critical in extracting the resistivity and power factor. We show that the sharp drop of resistance is attributed to the current that flows along the Si / $SiO_2$ interface for T > 200K through the accumulation layer of holes whose existence was confirmed by capacitance-voltage measurements and two band analysis of the Hall effect. However, we show that a large fraction of the current will flow along the top $Fe_3O_4$ at 300K giving a power factor of 70 mW/K²m when the $Fe_3O_4$ layer thickness is reduced down to 150 nm. In addition, we find that the heterostructures behave like p-n junctions in the transverse direction, which is responsible for the high thermopower.

**Experimental methods**

$Fe_{3-x}O_4$ films with different thicknesses of 150, 200, and 350 nm were grown by pulsed laser deposition (PLD) on *p*-type Si (001) at 210 $^o$C. The *p*-type Si (001) had a native $SiO_2$ thickness of 2 nm, which was verified by transmission electron microscopy. Details of the growth conditions and parameters are described elsewhere [13]. The films were analyzed regarding their phase purity, stoichiometry, crystallinity, preferred orientation growth, magnetic and electrical properties, which were compared with state-of-art magnetite films [13] High quality, polycrystalline films with the crystallites randomly oriented, x = 0.01 and uniform thicknesses of 150, 200, and 350 nm, as-determined by image analysis of their cross-sections observed in a scanning electron microscope, are considered in this article. Hall effect measurements were carried out in the Van der Pauw geometry between 100 and 300K using a current of I =10 $\mu A$ and a magnetic field of H = 6T. The transverse current-voltage (I-V) characteristics were measured between 100 and 300K after the formation of Au ohmic contacts on the $Fe_3O_4$ layer and the back side of the *p*-type Si (001) substrate as shown in the inset of Fig. 1. The capacitance - voltage (C-V) measurements were measured at 300K using the same structure shown as an inset in Fig. 1 and a frequency of 100 kHz.

The longitudinal current-voltage (I-V) characteristics were measured between 150 and 300K in a four probe configuration via the formation of Au contacts on the $Fe_3O_4$ layer as shown in Fig. 3 and then the resistivity was extracted by taking into account the thickness of $Fe_3O_4$. Finally, the in-plane Seebeck coefficient was determined by applying a thermal gradient on $Fe_3O_4$ and measuring the temperature and voltage differences between two points on the surface of $Fe_3O_4$ in the steady state as shown in Fig. 4. All of the above lateral and transverse current transport and thermopower measurements were carried out using a physical property measurement system (PPMS) by Quantum Design.

**Results and Discussion**

In order to understand in detail the transport and thermoelectric properties, it is necessary to consider the energy band diagram of the $Fe_3O_4$ / $SiO_2$ / *p*-type Si (001) heterostructure in thermodynamic equilibrium shown as an inset in Fig. 1. The *p*-type Si (001) substrate had a thickness of 500 $\mu m$ and a hole density of $10^{15}$ cm$^{-3}$ determined from Hall effect measurements at 300 K. From p ≈ $10^{15}$ cm$^{-3}$, we find that the Fermi level resides ≈0.25 eV above the valence band edge, i.e., $E_F - E_V$ = 0.25 eV. The work function of *p*-type Si(001) is given by $\varphi_S$ = + ($E_C$ - $E_F$) [22] i.e., $\varphi_S$ = 4.92 eV since the electron affinity $\chi_{Si}$ = 4.05 eV and

$E_C - E_F = 0.87$ eV. On the other hand, the work function of $Fe_3O_4$ is $\varphi_M = (5.5 \pm 0.05)$ eV [14]; so the work function difference between the metal and semiconductor $\varphi_M - \varphi_S = \varphi_{MS} = 0.58$ eV is positive, which is expected to produce an upwards band bending at the $Si/SiO_2$ interface leading to an accumulation of holes whose existence was confirmed by capacitance-voltage (C-V) measurements and by a two-band analysis of the Hall effect described in detail later.

Consequently, the n-$Fe_3O_4$ / $SiO_2$ / p-type Si(001) heterostructure is expected to act like a p-n junction and we find that it exhibits asymmetric I-Vs in the transverse direction as shown in Fig. 1. More specifically, the current shows an immediate increase for negatively applied voltages ($V_A$), i.e., negative potential on $Fe_3O_4$, but is initially zero for positive $V_A$ and then exhibits a substantial increase only for $V_A > + 2V$. This I-V characteristic is similar to that measured in a $Fe_3O_4$ / $SiO_2$ / n-type Si(001) heterostructure by Qu *et al*. [15] with an inversion layer of holes at the n-type $Si/SiO_2$ interface. Similarly, the 350 nm $Fe_3O_4$ / $SiO_2$ / p-type Si(001) heterostructure also exhibits asymmetric I-Vs as shown in Fig. 2, but while the current shows an immediate increase for negative $V_A$ we observe that $V_A > + 4V$ in order to establish a current flow. In both cases shown in Figs. 1 and 2, we find that the I-Vs depend on temperature, which was varied between 100 and 300 K, and also the thickness of $Fe_3O_4$. However, in order to understand the current transport we ought to consider the energy band diagrams under externally applied electric fields. Upon the application of an electric field, there will be an increase in band-bending at the Si / $SiO_2$ interface for negative potential on $Fe_3O_4$ as shown in Fig. 2. This will result into a further accumulation of holes at the Si / $SiO_2$ interface and spin down $t_{2g}$ electrons in the $Fe_3O_4$ will tunnel directly into the empty hole states resulting into a recombination current. In addition, spin down electrons residing in the $t_{2g}$ and $e_{2g}$ bands, where the $e_{2g}$ band is located $\approx 1.3$ eV above the $t_{2g}$ band [16], may cross the $SiO_2$ barrier into the *p*-type Si by thermally assisted tunneling. The I-Vs for negative $V_A$ follow a simple exponential form as shown in the inset of Fig. 2 and the diode may be thought as being forward biased. On the other hand, when the potential is positive on $Fe_3O_4$ the externally applied electric field will flatten out the potential as shown in Fig. 2 leading to a depletion of holes from the Si / $SiO_2$ interface and the current will be initially zero. A further, positive, increase of VA will attract electrons towards the Si / $SiO_2$ interface causing inversion and the current will then flow from the conduction band of Si(001) into the $Fe_3O_4$ so the p-n diode may then be thought as being reverse biased. A larger, positive voltage is required to drive a current perpendicularly through the $Fe_3O_4$ / $SiO_2$ / *p*-type Si(001) structure for larger thickness of $Fe_3O_4$. This may be attributed to an increase in the series resistance of the $Fe_3O_4$ layer since a change in the overall band line up and barrier height is not likely to occur by increasing the thickness of $Fe_3O_4$. In essence, the resistance of $Fe_3O_4$ is in series with the p-n heterojunction so a larger potential difference will appear across the $Fe_3O_4$ upon increasing its thickness. Similarly, a reduction in temperature will lead to a gradual suppression of thermionic assisted tunneling over the $SiO_2$ barrier but also a decrease of the carrier density and an increase in the resistance of the $Fe_3O_4$ layer since charge ordering should open a gap at the Fermi energy resulting into the metal-insulator transition below $T_V$ = 120 K [17] and a reduction in the current for a specific applied voltage. From the above, we conclude that the $Fe_3O_4$ / $SiO_2$ / *p*-type Si(001) heterostructure acts like a p-n junction, which is expected to limit the ohmic return current that occurs in bulk materials during charge redistribution under a temperature gradient resulting into higher thermopower as has been shown in p-n thermal diodes [18]. However, before elaborating further on the thermoelectric properties it is necessary to consider in more detail the current flow distribution in the lateral direction since this is critical in extracting the resistivity and power factor of the $Fe_3O_4$ / $SiO_2$ / *p*-type Si(001) devices.

In contrast to the I-Vs shown in Figs. 1 and 2, we find that the longitudinal I-Vs are linear-like, as shown in Fig. 3, with a slight non-linearity around the origin, giving a resistance of R = 5 kΩ for the 350 nm $Fe_3O_4$ and R = 8 kΩ for the 150 nm $Fe_3O_4$ heterostructure at 300 K. It appears at first sight that the longitudinal resistance is governed by the thickness of the $Fe_3O_4$. We have used noble metal contacts on $Fe_3O_4$, which are metallic-like in a four probe configuration as shown in Fig. 3 so the slight non-linearity is attributed to parallel conduction in the $Fe_3O_4$ and along the accumulation layer of holes at the Si / $SiO_2$ interface not due to the formation of a Schottky barrier between Au and $Fe_3O_4$.

In addition, the temperature dependence of the longitudinal resistance exhibits a sharp drop with increasing temperature as shown in the inset of Fig. 4. Similar temperature dependence has been reported in $Fe_3O_4$ / $SiO_2$ / n-type Si [19], $Fe_3C$ / $SiO_2$ / n-type Si [20], and other M / $SiO_2$ / Si structures, where M stands for very-thin metallic films [21] but these have been investigated as spintronic not thermoelectric devices. The sharp drop in resistivity has been attributed to the onset of conduction in the Si inversion layer of holes at the Si / $SiO_2$ interface, which gives a positive Hall voltage despite the fact that the substrate is n-type Si(001) [19–21]. More importantly, it has been shown that the current flows along the Si / $SiO_2$ interface in these M / $SiO_2$ / Si structures not through the bulk Si [21]. The existence of the accumulation layer of holes at the Si / $SiO_2$ interface of the $Fe_3O_4$ / $SiO_2$ / *p*-type Si heterostructure in our case was confirmed by capacitance voltage (C-V) measurements for the 200 nm $Fe_3O_4$ / $SiO_2$ / *p*-type Si(100) heterostructure as shown in the inset of Fig. 3, which is typical of a MOS capacitor with *p*-type Si [22]. For $V_A < 0$, holes are attracted to the Si / $SiO_2$ interface and the capacitance is essentially governed by the oxide thickness $C_{ox}$. For $V_A > 0$, holes are depleted from the interface and the capacitance starts decreasing due to the increasing capacitance of the depletion layer, which is in series with $C_{ox}$. The C-V curve of Fig. 3 is shifted positively by $\approx +0.25V$ suggesting the existence of an accumulation layer of holes at the Si / $SiO_2$ interface. This is further corroborated by the Hall voltage of the $Fe_3O_4$ / $SiO_2$ / *p*-type Si(100) heterostructure, which changed sign from being negative to positive around 220K and its magnitude was small and equal to + 0.02 mV at 300K and 6 T [12]. This gives a Hall carrier density of $4 \times 10^{16}$ $cm^{-3}$ assuming that it corresponds only to holes in the 500 *μ*m bulk *p*-type Si(100), which is larger than $10^{15}$ $cm^{-3}$ actually measured in *p*-type Si(001) only. Note that the Hall voltage for an electron density of $10^{21}$ $cm^{-3}$ typical for $Fe_3O_4$ [16] at 6T is equal to - 0.2 mV assuming one has a single layer of $Fe_3O_4$ with a thickness of 200 nm. These discrepancies exist as a result of parallel - conduction since the Hall coefficient depends on the relative magnitudes of the electron and hole carrier densities and mobilities according to $R_H = (p - nb^2) / e(p + nb)^2$, where $b = \mu_n / \mu_p$. Since $b = \mu_n / \mu_p = 0.5 / 500 \approx 10^{-3}$ and $R_H$ is positive one requires $p > nb^2$. Given that $n \approx 10^{21}$ $cm^{-3}$ at 300K this implies $p > 10^{15}$ $cm^{-3}$. If we take into account electrons in the $Fe_3O_4$ layer and holes at the Si / $SiO_2$ interface, we have $R_H = 0.7 \times 10^{-7}$ $m^3/C$ at 6T for 10 *μ*A taking a thickness of 200nm and Hall voltage of + 0.02 mV. Upon solving $R_H = (p - nb^2) / e(p + nb)^2$, we find that $p = 10^{19}$ $cm^{-3}$. On the other hand, if we consider that electrons flow along the $Fe_3O_4$ layer and the Si/$SiO_2$ interface but also along the *p*-type Si(001) bulk within 50 *μ*m of the Si / $SiO_2$ interface, we obtain $R_H = 0.17 \times 10^{-4}$ $m^3/C$ at 6T and 10 *μ*A for the same Hall voltage of t +0.02mV but $R_H = (p - nb^2) / e(p + nb)^2$ cannot be solved for p since there is no real solution, which is also the case for a thickness of 10 *μ*m keeping all else equal. Nevertheless, real solutions exist for thicknesses <1 *μ*m, which gives large hole densities of the order of $10^{19}$ $cm^{-3}$.

Now the sheet resistance and resistivity of the two dimensional hole gas (2DHG) with a density of $10^{19}$ $cm^{-3}$ and corresponding mobility of 100 $cm^2/Vs$ taking a thickness of 10 nm, which is typical of the spatial extent of the 2DHG in the quasi triangular potential well at the Si / $SiO_2$ interface [24] is 6 kΩ / □ and 6 mΩ cm, respectively. On the other hand, the resistivity of polycrystalline $Fe_3O_4$ grown on MgO by molecular beam epitaxy was found to decrease for thicknesses < 50nm according to Eerenstein *et al*. [25] but the resistivity did not vary significantly for $Fe_3O_4$ thicknesses between 50 and 100 nm and had a value of $\approx 5 m\Omega$ cm, which is close to the room temperature resistivity of the $Fe_3O_4$ / $SiO_2$ / *p*-type Si(100) structures shown in the inset of Fig. 4. A similar type of dependence has also been found in the case of $Fe_3O_4$ layers grown by Li et al. using PLD [26]. The low resistivity is a direct consequence of the large density of electrons, which is of the order of $10^{21}$ $cm^{-3}$ despite the low electron mobility as measured for instance by Reisinger *et al.* [27], who found a carrier density of 3 x$10^{21}$ $cm^{-3}$ in a 50 nm $Fe_3O_4$ film on MgO at 300K and a mobility of 0.48 $cm^2$/Vs. Consequently, the sheet resistance and resistivity of a $Fe_3O_4$ layer with a thickness of 150 nm taking a density of $10^{21}$ $cm^{-3}$ and mobility of 0.5 $cm^2$/Vs is 0.8 $k\Omega$ / □ and 12 $m\Omega$ cm, respectively. Note that the resistivities of the 2DHG and $Fe_3O_4$ layer are similar to those measured at 300K as shown in the inset of Fig. 4. However, the total resistance ($R_T$) of the $Fe_3O_4$ / $SiO_2$ / Si heterostructure is not only given by the resistance of the $Fe_3O_4$ layer ($R_F$) in parallel with the resistance of the 2DHG (R2DHG) at the Si/$SiO_2$ interface but also depends on the resistance of the $SiO_2$ barrier ($R_{OX}$), i.e., $R_T = R_F$ // ($2R_{OX} + R_{2DHG}$) as shown in the inset of Fig. 3. We do not expect a contribution from the *p*-type Si(001) bulk in accordance with Dai *et al*. [21], who showed that conduction occurs along the Si / $SiO_2$ interface in Cu / $SiO_2$ / Si heterostructures not through bulk Si. From the above, we conclude that most of the current will flow through the $Fe_3O_4$ layer since its sheet resistance is smaller than that of the 2DHG but also in view of the intervening $SiO_2$. It is worthwhile pointing out that the resistance of $Fe_3O_4$ alone decreases monotonically with temperature from several $M\Omega$ down to a few $k\Omega$ as shown in the inset of Fig. 3 and is close to the resistance of the 150nm $Fe_3O_4$ / $SiO_2$ / *p*-type Si(001) heterostructure at 300 K. Consequently, the sharp drop of the resistivity with temperature shown as an inset in Fig. 4 occurs due to the onset of tunneling of electrons from the $Fe_3O_4$ into the accumulation layer of holes at the Si / $SiO_2$ interface but conduction through the $Fe_3O_4$ layer is significant at 300 K. We ought to emphasize that we have observed this sharp drop of resistivity at the same temperature in Ni / $SiO_2$ / *p*-type Si(001) and Cu / $SiO_2$ / *p*-type Si(001) heterostructures similar to Dai et al [21], so it is not associated with the Verwey transition of $Fe_3O_4$. The sharp drop of resistance with increasing temperature is smaller and less abrupt, i.e., broader, by increasing the thickness of the $Fe_3O_4$ layer suggesting that a larger fraction of current flows through the $Fe_3O_4$ since its resistance will be smaller resulting into a smaller electric field perpendicular to the Si / $SiO_2$ interface that will drive current through the $SiO_2$ barrier and along the Si / $SiO_2$ interface.

While the resistivity exhibited a sharp drop for T > 200 K, we find that the thermopower measured on top of the $Fe_3O_4$ increased over the same temperature range as shown in Fig. 4, something that would not be observed in bulk materials where the carrier density must be reduced in order to increase the thermopower. The thermopower of single crystal magnetite [12, 23] is negative over the entire measured temperature range, i.e., S = - 60 $\mu$V/K. On the other hand, the thermopower of the $Fe_3O_4$ / $SiO_2$ / *p*-type Si(001) structure is small and positive, remains constant up to 200K and then exhibits a large increase reaching quite large values ≈ + 1000 $\mu$V/K at 300 K. It is useful to note that the thermopower of *p*-type Si(001) without the $Fe_3O_4$ layer was considerably higher than + 200 $\mu$V/K between 150 and 200K and reached + 900 $\mu$V/K at 300 K, which is comparable to the thermopower measured in the case of $Fe_3O_4$ / $SiO_2$ / *p*-type Si(001). In addition, it is useful to point out that the thermopower decreased significantly upon increasing the thickness of the $SiO_2$ so clearly the large thermopower is not due to the $Fe_3O_4$ or $SiO_2$, which is an insulator that is void of carriers and has a large energy band gap of 9 eV. The large thermopower is clearly related to the charge redistribution of holes, which occurs in the underlying *p*-type Si(001). More specifically, the application of a temperature gradient along the surface will result into a transverse temperature gradient so electrons on the high temperature side will cross the barrier into the underlying Si(001), which is at a lower temperature and the quasi Fermi level in the semiconductor will be lower than that in the metallic $Fe_3O_4$ similar to the energy band diagram shown as an inset in Fig. 2. At low temperatures, only a limited number of carriers will undergo tunneling from the $Fe_3O_4$ into the Si(001) but for temperatures above 200K a larger number of electrons will cross into the underlying *p*-type Si(001), which in turn allows one to measure the potential difference that develops as a result of the charge redistribution in the *p*-type Si(001) due to the applied temperature gradient. We should note that hot carrier filtering through barriers with heights of many kT [28] or carrier injection using p-n thermal diodes [18] have been shown to yield high S and ZT although the precise physical mechanisms are not yet well understood. It is, therefore, possible that hot carrier filtering occurs due to the specific electronic structure of $Fe_3O_4$ and the existence of the two spin down bands as shown in Fig. 2, while the formation of a p-n like junction will limit the ohmic return current which occurs upon charge redistribution under a temperature gradient.

Evidently, the current transport and thermoelectric properties of the $Fe_3O_4$ / $SiO_2$ / *p*-type Si(001) heterostructure may be exploited for the realization of high thermoelectric figure of-merit devices by reducing j and exploiting its high thermoelectric power factor. Silicon nanowires exhibit a one-hundred fold reduction in thermal conductivity compared to the bulk giving very high ZT [9] so we anticipate that $Fe_3O_4$ / $SiO_2$ / Si core-shell nanowires are very promising for the realization of emerging devices in future integrated circuits for power generation or cooling. However, the origin of the high thermopower, thermal conductivity, and thermoelectric figure of merit will be presented in more detail elsewhere.

**Conclusions**

In conclusion, we have investigated the lateral and transverse current transport and thermoelectric properties of $Fe_3O_4$ / $SiO_2$ / *p*-type Si(001) heterostructures with $Fe_3O_4$ thicknesses of 150, 200, and 350 nm between 100 and 300 K. We show that the sharp drop of the in-plane resistance is attributed to the onset of conduction along the Si / $SiO_2$ interface for T > 200K due to tunneling of electrons from the $Fe_3O_4$ into the accumulation layer of holes at the Si / $SiO_2$ interface whose existence was confirmed by capacitance - voltage measurements and two band analysis of the Hall effect giving a hole density of the order of $10^{19}$ $cm^{-3}$ and a sheet resistance of 6 k$\Omega$ / □ at 300 K. In contrast, the resistance of $Fe_3O_4$ on its own, decreases monotonically with increasing temperature without exhibiting any sharp drop for T > 200K reaching 0.8 k$\Omega$ / □ at 300 K. The sharp drop of the in-plane resistance is accompanied by a large increase of the Seebeck coefficient over the same temperature range reaching +1000 $\mu$V/K but a large fraction of the current will flow along the top $Fe_3O_4$ layer at 300K giving a power factor of 70 mW/$K^2$m when the $Fe_3O_4$ layer thickness is reduced down to 150 nm. Finally, we show that the $Fe_3O_4$ / $SiO_2$ / *p*-type Si(001) heterostructures exhibit asymmetric current-voltage characteristics in the transverse direction and behave like p-n junctions, which is responsible for the high thermopower by preventing the ohmic return current upon charge redistribution under a temperature gradient.

**Acknowledgements**

This work was supported in part by the Cyprus Research Promotion Foundation (Project ANAVATHMISI/ 0609/06) and by the Portuguese Foundation for Science and Technology grant to ICEMS.

**Figure Captions**

**FIG. 1**. I-V characteristics in the transverse direction for the 150nm $Fe_3O_4$ / $SiO_2$ / $p$-type Si(001) structure between 100 and 300 K. Inset shows the energy band diagram in thermodynamic equilibrium and a schematic of the device and measurement.

**FIG. 2.** I-V characteristics in the transverse direction for the 350 nm $Fe_3O_4$ / $SiO_2$/ $p$-type Si(100) structure between 100 and 300K. Top left schematic shows the energy band diagram for $V_A < 0$; right schematic for $V_A > 0$. The lower inset shows a simple exponential fit to the I-V characteristics.

**FIG. 3.** Longitudinal I-V characteristics for the 150nm $Fe_3O_4$ / $SiO_2$ / $p$-type Si(100) structure between 100 and 300 K. Top left inset shows corresponding C-V curve obtained at T = 300K and 100 kHz. Lower right inset shows the temperature dependence of the resistance of $Fe_3O_4$ on glass and 20 nm $SiO_2$ / $p$-type Si(001) along with the temperature dependence of the $Fe_3O_4$ / $SiO_2$ / $p$-type Si(100) heterostructure including the native oxide.

**FIG. 4.** Temperature dependence of thermopower for the $Fe_3O_4$ / $SiO_2$ / $p$-type Si(100) structure between 100 and 300K and different thicknesses of $Fe_3O_4$. Insets show the corresponding temperature dependence of the resistivity and relevant schematic of the device for the measurement of thermopower.

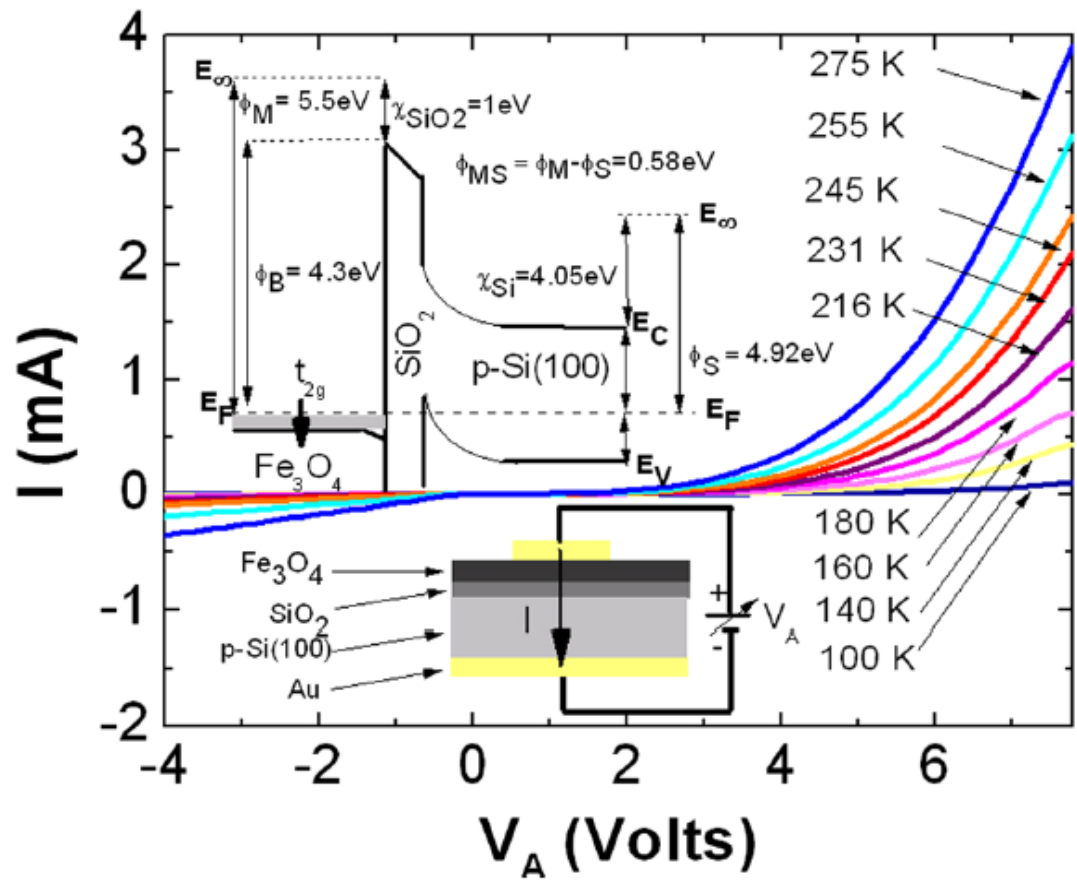

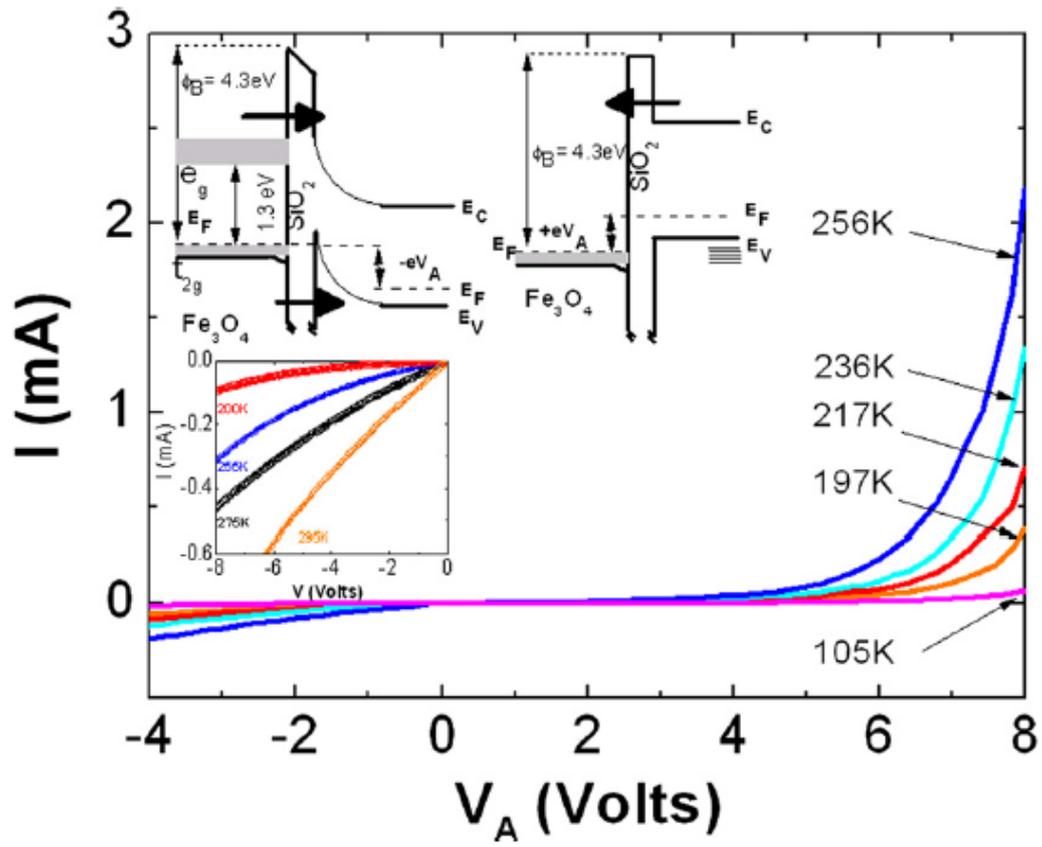

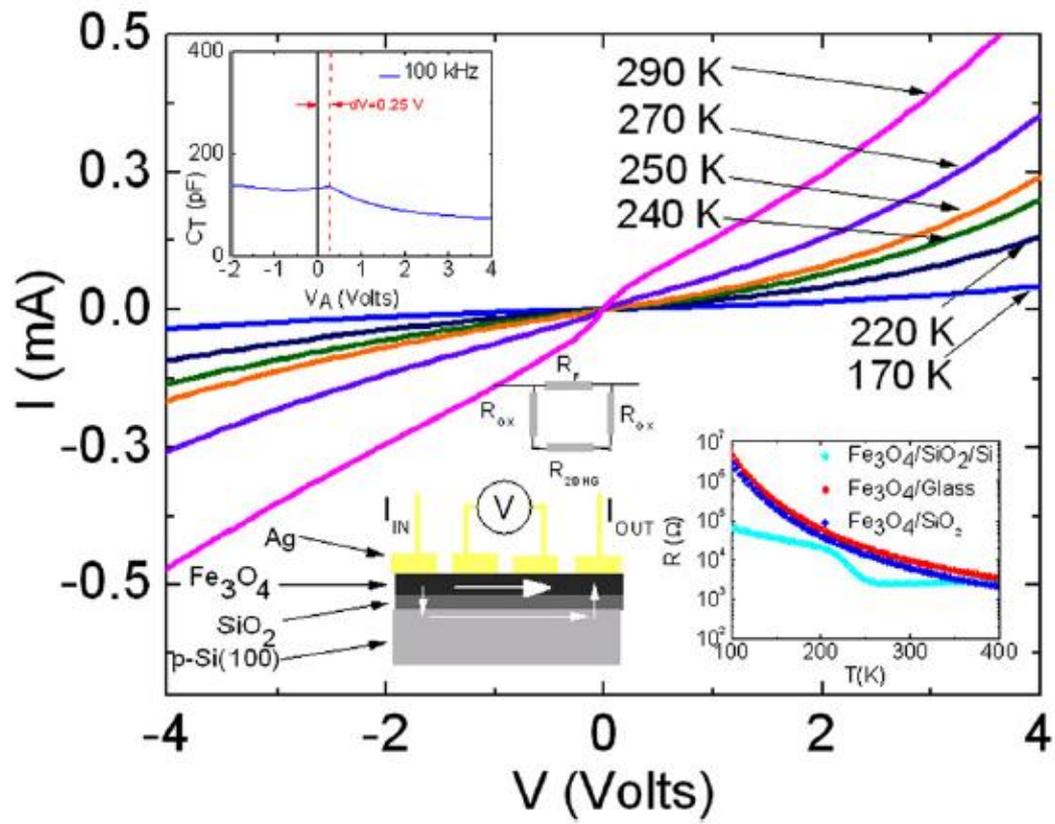

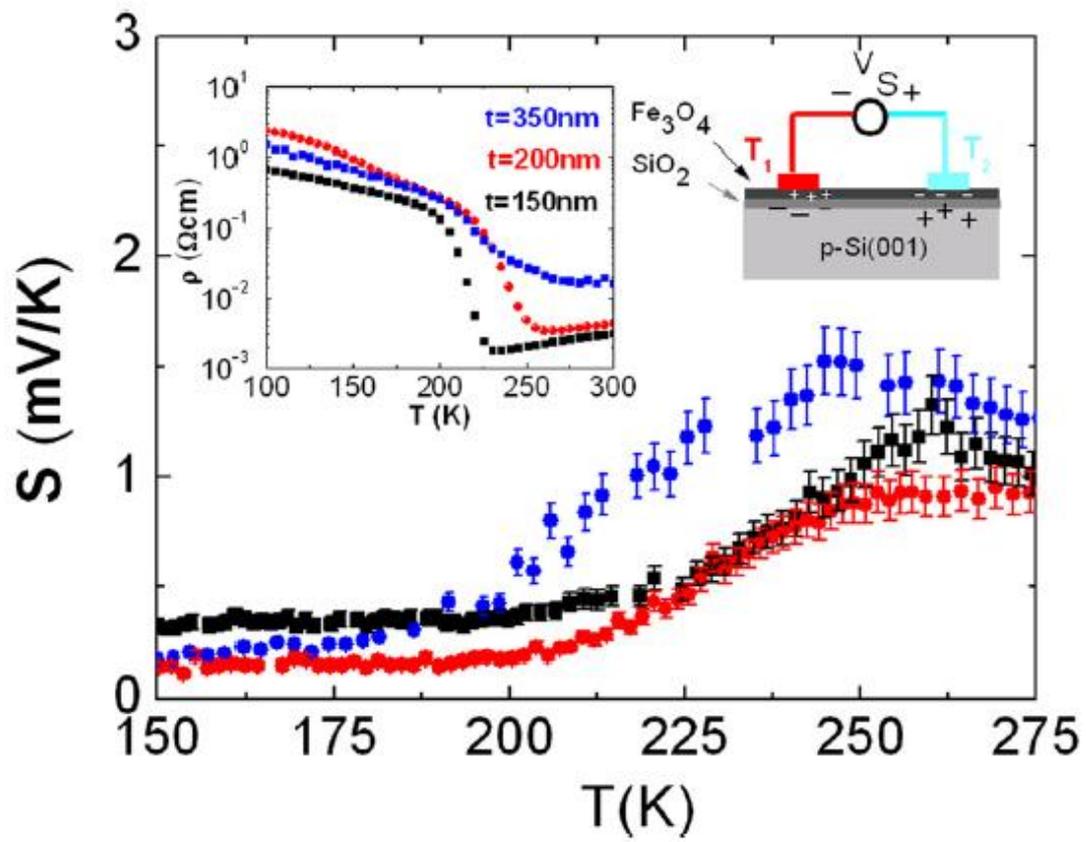